\documentclass[aps,prd,twocolumn,showkeys,floats,showpacs, nofootinbib]{revtex4-1}
\usepackage{graphics,epsfig,color} 
\usepackage{amsmath,amssymb}
\usepackage{hyperref}

\def\d {\mbox{d}}
\newcommand{\ul}[1]{\overline{#1}}
\newcommand{\ut}[1]{\tilde{#1}}

\def\be{\begin{equation}}
\def\ee{\end{equation}}
\def\bea{\begin{eqnarray}}
\def\eea{\end{eqnarray}}
\begin{document}
\title{The $\gamma$ parameter  in Brans-Dicke-like (light-)Scalar-Tensor theory with a universal scalar/matter coupling}
\author{Olivier Minazzoli}
\affiliation{UMR ARTEMIS, CNRS, University of Nice Sophia-Antipolis,
Observatoire de la C\^ote d’Azur, BP4229, 06304, Nice Cedex 4, France}
\affiliation{Jet Propulsion Laboratory, California Institute of Technology,\\
4800 Oak Grove Drive, Pasadena, CA 91109-0899, USA}
\begin{abstract}
The post-Newtonian parameter $\gamma$ resulting from a universal scalar/matter coupling is investigated in Brans-Dicke-like Scalar-Tensor theories where the scalar potential is assumed to be negligible. Conversely to previous studies, we use a perfect fluid formalism in order to get the explicit scalar-field equation. It is shown that the metric can be put in its standard post-Newtonian form. However, it is pointed out that $1-\gamma$ could be either positive, null or negative for finite value of $\omega_0$, depending on the coupling function; while Scalar-Tensor theories without coupling always predict $\gamma<1$ for finite value of $\omega_0$. 
\end{abstract}
\maketitle


\section{Introduction}
Brans-Dicke-like scalar-tensor theories are known to be good alternative candidates to General Relativity (GR) \cite{GEF2004,Will_book93,Will-lrr-2006-3,damourCQG92}. Similar theories with both scalar/curvature and scalar/matter couplings generically appear in (gravitational) Kaluza-Klein theories with compactified dimensions \cite{Sstring88,fujiiBOOKst}, or in string theories at the low energy limit \cite{DamPolyNPB94,damourPRD96,damourPRL02,gasperiniPRD02}. From a more phenomenological point of view, it seems that some restrictions, such as gauge and diffeomorphism invariance, single out such type of theories as well \cite{armendarizPRD02}. Recently, scalar/matter couplings have been introduced in several different type of theories:  in $f(R)$ gravity \cite{nojiriPhLB04,bertolamiPRD07,bertolamiPRD08,bertolamiCQG08,sotiriouCQG08,defeliceLLR13,sotiriouRMP10,nojiriPhR11,harkoPRD13}, in Brans-Dicke theories \cite{dasPRD08,bisabrPRD12,moffatIJMPD12}, or in the so-called MOG (MOdified Gravity) \cite{moffatJCAP06,moffatCQG09}.
Such theories are often invoked as a possible explanation for dark Energy -- which is generically attributed to a scalar field \cite{peebleRMP03} --, for the possible observed variation of the fine structure constant in both time \cite{webbPRL2001} and space \cite{webbPRL11,murphyMNRAS03}, for (at least) some phenomena usually attributed to dark matter; or to generically predict violations of the equivalence principle \cite{DamPolyNPB94,dvaliPRL02,olivePRD08,damourPRD10,damourCQG12}.

The post-Newtonian phenomenology of theories with scalar/matter coupling has partially been studied, notably in \cite{DamPolyGRG94,DamPolyNPB94,damourPRD10}. These studies concentrate on a possible dynamical decoupling mechanism that would allow theories with a scalar/matter coupling to pass Solar system tests on the various versions of the equivalence Principle. Among other results, they find that after decoupling, the post-Newtonian parameter of such theories has to be very close to one; but is always less than one, just as in regular Brans-Dicke scalar-tensor theories. Here we demonstrate with a very simple calculation that the very last statement is not true in general: depending on the scalar/matter coupling function, the post-Newtonian parameter can be less, equal or more than one. 

Our result is based on a perfect fluid approach while the mentioned previous studies are based on a non-interacting particles approach. After re-writing our equations in the so-called Einstein representation, we demonstrate that the difference between \cite{DamPolyGRG94,DamPolyNPB94} and this paper is not due to the different formalisms used (non-interactive point particles versus perfect fluid);  but rather comes from on a wrong property assumed in \cite{DamPolyGRG94,DamPolyNPB94}.

In section \ref{sec:eqm}, we derive the equations of motion coming from the considered action. Then, in section \ref{sec:magnitude} we concentrate on the post-Newtonian parameter $\gamma$ resulting from such theories. In section \ref{sec:string} we make the connection with the results presented in \cite{DamPolyGRG94,DamPolyNPB94}. Finally, we give our conclusions in \ref{sec:concl}. 

\section{Equations of motion}
\label{sec:eqm}

The action describing Brans-Dicke-like theories with a universal scalar/matter coupling can be written as follows:
\begin{equation}\label{eq:action}
S=\int d^4x\sqrt{-g} \left( \Phi R -
\frac{\omega(\Phi)}{\Phi} \left(\partial_{\sigma}\Phi \right)^2+ 2f(\Phi)\mathcal{L}_m(g_{\mu \nu},\Psi) \right),
\end{equation}
where $g$ is the metric determinant, $R$ is the Ricci scalar
constructed from the metric $g_{\mu \nu}$ , $\mathcal{L}_m$ is the material Lagrangian and $\Psi$ represents the non-gravitational fields. From this action, and defining
\begin{equation}
T_{\mu \nu}=-\frac{2}{\sqrt{-g}} \frac{\delta(\sqrt{-g}\mathcal{L}_m)}{\delta g^{\mu \nu}},
\end{equation}
 one gets the following equations of motion:
\begin{eqnarray}
R_{\mu \nu}-\frac{1}{2}g_{\mu \nu}R&=& \frac{f(\Phi)}{\Phi}T_{\mu
\nu}+\frac{\omega(\Phi)}{\Phi^2}(\partial_{\mu} \Phi \partial_{\nu} \Phi
- \frac{1}{2}g_{\mu \nu}(\partial_{\alpha}\Phi)^2) \nonumber \\
&& + \frac{1}{\Phi} [\nabla_{\mu} \nabla_{\nu} -g_{\mu \nu}\Box]\Phi ,\label {eq:motiong}
\end{eqnarray}
and
\begin{equation}\label{eq:motionPhi}
\frac{2\omega(\Phi)+3}{\Phi}\Box \Phi= \frac{f(\Phi)}{\Phi} T - 2 f_{,\Phi}(\Phi) \mathcal{L}_m - \frac{\omega_{,\Phi}(\Phi)}{\Phi} (\partial_\sigma \Phi)^2 .
\end{equation}

\section{The parameter $\gamma$}
\label{sec:magnitude}

In this section, we are interested in showing that the parameter $\gamma$ can take different values than usually expected. Therefore we develop the equations at the $c^{-2}$ level only. Let us write the perturbations of the fields  as follow:
\begin{eqnarray}
&&\Phi=\Phi_0+c^{-2} \varphi \label{eq:dev_phi}\\ 
&&g_{\mu \nu}=\eta_{\mu \nu}+ c^{-2} h_{\mu \nu}+O(c^{-3})\label{eq:dev_g},
\end{eqnarray}
where $\eta_{\mu \nu}$ is the metric of Minkowki and $\Phi_0$ is constant background field \footnote{For basic principles about post-Newtonian developments, see for instance \cite{damourBOOK87,DSX-I,Kopeikin_Vlasov_2004} and references therein. In \ref{sec:devSF}, we recall the reason why the perturbation of the scalar field can be developed using the same small parameter as with the metric.}.  Now, if one assumes the conservation of the matter fluid current ($\nabla_\sigma(\rho U^\sigma)=0$, where $c^2 \rho$ is the \textit{rest mass energy density} and $U^\alpha$ the four-velocity of the fluid), one has $\mathcal{L}_m=-\epsilon$, where $\epsilon$ is the \textit{total energy density} \cite{fockBOOK64,brownCQG93,bertolamiPRD08,harkoPRD10,moi_hPRD12,moiPRD13}. Therefore, at the first order in the post-Newtonian development, one has $\mathcal{L}_m=-c^2 \rho + O(c^0)=T+O(c^0)$. Hence, equations (\ref{eq:motiong}) and (\ref{eq:motionPhi}) can be re-written at the first perturbative order as follows: 
\begin{eqnarray}
&&R^{\mu \nu}=\frac{f(\Phi_0)}{\Phi_0} \left(T^{\mu \nu}-\frac{1}{2} g^{\mu \nu} T \right)  \label{eq:Rmunu1} \\
&&~~~~~~~+ \frac{1}{\Phi_0} \left(\partial^\mu \partial^\nu+ \frac{1}{2} g^{\mu \nu} \triangle \right) \Phi+O(c^{-3}),\notag \\
&&\frac{2\omega +3}{\Phi_0} \Box \Phi = \left(1+\Upsilon \right)\frac{f(\Phi_0)}{\Phi_0} T+O(c^{-3}),
\end{eqnarray}
where
\begin{equation}
\Upsilon \equiv  - 2 ~\Phi_0~ \frac{\partial \ln{f(\Phi)}}{\partial \Phi} |_{\Phi_0}.
\end{equation}
Defining 
\begin{eqnarray}
\sigma &\equiv & T^{00}/c^2+O(c^{-2}),\\
 G_{eff} &\equiv & \left(1 + \frac{1+\Upsilon}{2 \omega_0 +3} \right) \frac{c^4}{8 \pi} \frac{f(\Phi_0)}{\Phi_0},\\
 \gamma &\equiv & \frac{2 \omega_0 +2 - \Upsilon}{2 \omega_0 + 4 + \Upsilon},\label{eq:gammadef}
 \end{eqnarray}
where $\omega_0=\omega(\Phi_0)$, the previous equations can be re-written as follows:
 \begin{eqnarray}
 &&R^{00}=c^{-2} \left\{ 4 \pi G_{eff} \sigma \right\}+O(c^{-3}),\\
 &&R^{ij}=c^{-2} \left\{-\delta_{ij} \gamma 4 \pi G_{eff} \sigma + \frac{1}{\Phi_0} \partial_i \partial_j \varphi \right\}+O(c^{-3})\\
 &&\frac{1}{\Phi_0} \triangle \varphi = - \frac{2+2\Upsilon}{2 \omega_0 + 4 + \Upsilon}  4 \pi G_{eff} \sigma+O(c^{-1}). \label{eq:phi2omeg}
 \end{eqnarray} 
 It is then straightforward to show that the metric solution can be put under the following standard post-Newtonian form:
 \begin{eqnarray}
 &&g_{00}=-1+c^{-2}\frac{2 w}{c^2}+O(c^{-3}),\label{eq:main_resulti}\\
 &&g_{0i}=O(c^{-3}),\\
 &&g_{ij}=\delta_{ij}\left(1+c^{-2}\frac{2 \gamma w}{c^2} \right)+O(c^{-3}),
 \end{eqnarray}
 where $\gamma$ is indeed a constant given by (\ref{eq:gammadef}), and where $w$ satisfies the equation of Newton at the first perturbative order :
 \begin{equation}
 \triangle w = -4 \pi G_{eff} \sigma +O(c^{-1}).\label{eq:main_resulte}
 \end{equation}
 The important fact to notice is that, depending on the value of $\Upsilon$ (and thus depending on the coupling function), $1-\gamma$ could be either positive, null \footnote{See also the paper of Moffat and Toth \cite{moffatARXIV10} in which they explored such a possibility in order to argue the possible solar system viability of Modified Gravity Theory (MOG) \cite{moffatJCAP06,moffatCQG09}.} or negative; while STT without coupling predict a positive value for finite value of $\omega_0$. In particular, let us notice that $f(\Phi) \propto \Phi$ (such as for the low energy action of string theories at tree-level \cite{DamPolyGRG94}) leads to $\gamma>1$. Otherwise, it is also interesting to note that $f(\Phi) \propto \sqrt{\Phi}$ implies $\gamma=1$ \cite{moiPRD13_2} \footnote{One can show that it implies $\beta=1$ as well \cite{moiPRD13_2}.}.

\section{The string dilaton case}
\label{sec:string}

 Let us remind that the action (\ref{eq:action}) is a generalization of the low-energy action predicted by string theories at tree-level (see equation (1) in \cite{DamPolyGRG94}); and a special case of the assumed action after full string loop expansion (see the second action in \cite{DamPolyNPB94}).  

As one can see, \cite{DamPolyGRG94,DamPolyNPB94} do not predict that the post-Newtonian constant $\gamma$ could be more than, or equal to one; while the main result of the present paper is to show that the scalar/matter coupling implies that $\gamma$ could be exactly equal to one, or be either less or more than one -- depending on the coupling function $f(\Phi)$. However, \cite{DamPolyGRG94,DamPolyNPB94} based their results on a non-interactive point-particle formalism; while the present paper is based on a perfect fluid formalism. Therefore, at a first glance, it seems that the two formalisms lead to different results. 

However, it turns out that \cite{DamPolyGRG94,DamPolyNPB94} have a mistake in the definition of a coupling parameter. This mistake leads to the apparent difference of results between the two formalisms. Indeed, as matter of fact, the two formalisms are equivalent for pressure-less fluids \cite{moiPRD13} and therefore one should expect that the two formalisms predict the same outcome in this limit.

\cite{DamPolyGRG94,DamPolyNPB94} work in the Einstein representation (also known as the Einstein \textit{frame}) such that, with the notations of the present paper, their action writes:
\bea
S_{DamPoly}&=& \int d^4x \sqrt{-\ut{g}} \left(\frac{1}{4q} \tilde{R} - \frac{1}{2q} (\nabla \ul{\varphi})^2 \right) \notag \\
&&- \sum_\textrm{particles} \int \tilde{m}(\ul{\varphi})c d\tilde{s},
\eea
where $q$ is a coupling constant, $\ut{m}$ is the mass of particles in the Einstein representation and $\ut{g}_{\alpha \beta}$ is the metric in the Einstein representation -- related to the original representation by the conformal scalar $B_g$ through $\ut{g}_{\alpha \beta}=C B_g ~g_{\alpha \beta}$, where $C$ is some numerical constant. The resulting equations of motion write:
\bea
\ut{R}_{\mu \nu}= 2 \partial_\mu \ul{\varphi} \partial_\nu \ul{\varphi}+2q\left(\ut{T}_{\mu \nu}- \frac{1}{2} g_{\mu \nu}\ut{T} \right),\label{eq:Reinstein}\\
\ut{\Box} \ut{\varphi}= -q  \alpha \ut{T}, \label{eq:faussescalar}
\eea
where $\alpha$ is defined in \cite{DamPolyGRG94,DamPolyNPB94} as $\alpha=\partial \ln \ut{m} / \partial \ul{\varphi}$ and where we considered only one gravitational source (one particle)  in order to simplify the notations \footnote{Indeed, the paper deals with notations introduced in more than two papers and might become unnecessary difficult to follow without this simplification -- that does not change the discussion otherwise.}. Then, \cite{DamPolyGRG94,DamPolyNPB94} use an equation given in \cite{damourCQG92} that gives the parameter $\gamma$ as a function of the coupling parameter $\alpha$. The equation reads
\be
\gamma-1=- \frac{2 \alpha^2}{1+\alpha^2}|_{\Phi_0}, \label{eq:gammagiven}
\ee
such that $\gamma<1$ for finite real value of $\alpha |_{\Phi_0}$. The important point to notice is that in (2.7d) in \cite{damourCQG92}, $\alpha$ is defined as $\alpha=\partial \ln A / \partial \ul{\varphi}$ -- where $A$ is the square-root of the conformal factor given by $\ut{g}_{\alpha \beta}=A^{-2}(\ul{\varphi}) g_{\alpha \beta}$. Thus, identifying the definitions used in \cite{damourCQG92} and in \cite{DamPolyNPB94}, one has $C B_g = A^{-2}$ -- and identifying with the notations of the current paper, one has $\Phi=C B_g=A^{-2}$. On the other hand in \cite{DamPolyGRG94,DamPolyNPB94}, $\alpha$ is defined as $\alpha=\partial \ln \ut{m} / \partial \ul{\varphi}$, where $\ut{m}$ is the mass of the particle in the Einstein representation. In usual Brans-Dicke-like theories (ie. when $f(\Phi)$ is a constant) $\partial \ln A / \partial \ul{\varphi} = \partial \ln \ut{m} / \partial \ul{\varphi}$ since in that case one simply has $\ut{m} = A~ m$, where $m$ is the constant mass of the particle in the Jordan representation. However, the equality does not hold in the general case when $f(\Phi)$ is not a constant, and one has  $\partial \ln A / \partial \ul{\varphi} \neq \partial \ln \ut{m} / \partial \ul{\varphi}$ in general. Indeed, in general one has:
\bea
\frac{\partial \ln \ut{m}(A(\varphi),f(\varphi))}{\partial \varphi}&=&\frac{\partial \ln \ut{m}(A(\varphi),f(\varphi))}{\partial A} \frac{\partial A}{\partial \varphi}\label{eq:boom} \\
&&+\frac{\partial \ln \ut{m}(A(\varphi),f(\varphi))}{\partial f} \frac{\partial f}{\partial \varphi}.  \notag
\eea
And because $f(\varphi)$ is in general independent to $A(\varphi)$, the last terms in (\ref{eq:boom}) shows that $\partial \ln A / \partial \varphi \neq \partial \ln \ut{m} / \partial \varphi$ in general. Now in particular, let us notice that \cite{DamPolyNPB94} assume that
\be
\ut{m}= \mu B_g^{-1/2} e^{-8 \pi^2 \nu B_g} \Lambda, \label{eq:assumpt}
\ee
where $\mu$ and $\nu$ are pure number of the order of unity and $\Lambda$ is the string cut-off mass scale. Since one has $C B_g = A^{-2}$, one has $\partial \ln A / \partial \ul{\varphi} \neq \partial \ln \ut{m} / \partial \ul{\varphi}$. Therefore using equation (\ref{eq:gammagiven}) is not appropriate in the context considered by \cite{DamPolyGRG94,DamPolyNPB94,damourPRD10,damourPRL02,damourPRD02} -- even if the assumption (\ref{eq:assumpt}) was correct. 

Now, as in appendix \ref{sec:RSSF}, let us define $\alpha_0=\partial \ln A / \partial \ul{\varphi} |_{\Phi_0}$, and the coupling strength $\alpha_2$ by:
\be
\ut{\Box} \ul{\varphi}= -q  \alpha_2 \ut{T}. \label{eq:phiA2}
\ee
According to the previous discussion $\alpha_0 \neq \alpha_2$ in general. The conformal transformation of the Einstein metric to the metric in the original frame involves the transformation $g_{\alpha \beta}=A^2~\tilde{g}_{\alpha \beta}=[A_0^2+ 2 c^{-2} ~A_0 (\partial A/\partial \ul{\varphi})_0 ~\delta \ul{\varphi}+O(c^{-4})]~\tilde{g}_{\alpha \beta}$, where $\ul{\varphi}=\ul{\varphi}_0+c^{-2} \delta \ul{\varphi}$. Let us consider $A_0=1$ -- that simply means that one keeps the same metric's units in the two representations at the present epoch and does not restrict the generality -- one has $A_0(\partial A/\partial \ul{\varphi})_0 = \alpha_0$ and therefore $g_{\alpha \beta}= [1+ 2 c^{-2}~ \alpha_0 ~\delta \ul{\varphi}+O(c^{-4})]~\tilde{g}_{\alpha \beta}$.  Now, from equation (\ref{eq:Reinstein}), one deduces:
\be
c^{-2}\triangle \ut{w}= - q \ut{T}^{00} + O(c^{-4}),
\ee
where $\tilde{w}$ is the scalar potential of the Einstein metric. Therefore, from (\ref{eq:phiA2}), one deduces:
\be
\delta \ul{\varphi}= -  \alpha_2 \ut{w}+O(c^{-2}).
\ee
Hence one gets:
\be
g_{\alpha \beta}= \left[1-2c^{-2}~ \alpha_0 \alpha_2~ \tilde{w}+O(c^{-4})\right] ~\tilde{g}_{\alpha \beta}.
\ee
Developing the Einstein metric (that is such that it satisfies the so-called Strong Spatial Isotropic Condition (SSIC) -- ie. $\tilde{g}_{ij} \tilde{g}_{00} = - \delta_{ij} + O(c^{-4})$ \footnote{It has to be noticed that from (\ref{eq:Reinstein}) and (\ref{eq:faussescalar}), one gets $\ut{R}_{ij} - 1/2 \ut{g}_{ij} \ut{R}=O(c^{-4})$. Therefore, one can algebraically deduce that the Einstein metric can be expressed in a set of coordinates for which the metric satisfies the SSIC. For the derivation of this algebraic result, see \cite{DSX-I}.}), one gets the following equation for $\gamma$: 
\be
\gamma-1= - \frac{2 \alpha_0 \alpha_2}{1+ \alpha_0 \alpha_2}.
\ee
Hence, remembering that from solar system constraints one has $|\alpha_0 \alpha_2| \ll 1$, $\gamma-1$ can be positive if $\textrm{sign}(\alpha_0)=-\textrm{sign}(\alpha_2)$. Now, as demonstrated in appendix \ref{sec:RSSF}, $\alpha_2=(1+\Upsilon) \alpha_0$ and therefore the equation for $\gamma$ results to:
\be
\gamma-1= - \frac{2 (1+\Upsilon) \alpha_0^2}{1+ (1+\Upsilon)\alpha_0^2},
\ee
which corresponds to the result given by equation (\ref{eq:gammadef}) (because $\alpha_0^{-2}= 2 \omega_0+3$). In particular, one recovers the fact that $\gamma>1$ for $\Upsilon<-1$. Let us add that in the context of the universal coupling considered in \cite{DamPolyGRG94,DamPolyNPB94}, that is predicted by string theory at tree level, it means that $\gamma>1$; while \cite{DamPolyGRG94,DamPolyNPB94} say that $\gamma$ should be less than one as in usual scalar-tensor theories.

\section{Conclusion and final remarks}
\label{sec:concl}

In this paper we have shown that a universal scalar/matter coupling modifies the usual expression of the post-Newtonian parameter $\gamma$ in such a way that $1-\gamma$ could be either positive, null or negative for finite value of $\omega_0$; while it is usually thought to be positive only. In particular, we pointed out that previous studies considering similar couplings have missed that fact and we gave the reason for the apparent discrepancy.

\appendix 

\section{Development of the scalar field}
\label{sec:devSF}

The assumption that the scalar-field perturbation can be developed with the same small parameter as with the metric is justified by the field equations. Indeed, the sources of the two field equations are both proportional to the matter density: $\mathcal{L}_m \sim -c^2 \rho$ and $ \sim T \sim - T^{00} \sim -c^2 \rho$ -- the first equality is demonstrated in \cite{moi_hPRD12}; while the others come from the post-Newtonian assumptions as explained in \cite{DSX-I}. Therefore, unless $\Phi f_{,\Phi}/f$ is big, the relative perturbation of the scalar field is of the order of the relative perturbation of the metric. However, the PN parameter $\gamma$ is already measured to be very close to 1. Therefore, the relative perturbation of scalar field is necessarily much smaller than the relative perturbation of the metric and one does not have $\Phi f_{,\Phi}/f$  big in general -- at least in the solar system's neighborhood. 

Now since the order of magnitude of the relative perturbation of the scalar field is at best of the order of the relative perturbation of the metric, one can parametrize the development of the scalar field with the same parameter as with the metric. 

Let us stress that it is the usual procedure in post-Newtonian developments of alternative theories of gravitation (see, for instance, \cite{Will_book93,Kopeikin_Vlasov_2004,moiCQG12,dengPRD12}).

\section{Using the Einstein representation}
\label{sec:EF}

The results presented in this paper do not depend on the representation used to do the calculations. However, it is always interesting to re-derive the calculations in the Einstein representation in order to check the results obtained while using the original representation only. The action writes in the original and Einstein representation respectively as follows:
\begin{eqnarray}
S&=&\int d^4x\sqrt{-g} \left( \Phi R -
\frac{\omega(\Phi)}{\Phi} g^{\alpha \beta} \partial_\alpha \Phi \partial_\beta \Phi \right) + S_m,\\
&=& \int d^4x\sqrt{-\tilde{g}} \left(  \tilde{R} -
\left(\omega(\Phi(\varphi))+\frac{3}{2} \right) \tilde{g}^{\alpha \beta} \partial_\alpha \varphi \partial_\beta \varphi \right) + S_m, \nonumber \\ \label{eq:actionER}
\end{eqnarray}
where $g^{\alpha \beta} \equiv \Phi \tilde{g}^{\alpha \beta}$, $\sqrt{-g}= \Phi^{-2} \sqrt{-\tilde{g}} $ and $\varphi \equiv \ln \Phi$. By definition, the material part of the action ($S_m$) writes:
\begin{eqnarray}
S_m&=&\int d^4x\sqrt{-g}~ 2 f(\Phi) \mathcal{L}_m(g_{\mu \nu}, \Psi),\label{eq:actionMJ}\\
&=& \int d^4x\sqrt{-\tilde{g}} ~2 f(\Phi(\varphi)) \tilde{\mathcal{L}}_m(\ut{g}_{\nu \nu}, \Phi, \Psi).
\end{eqnarray}
Therefore, by definition, one has $\tilde{\mathcal{L}}_m=\Phi^{-2}\mathcal{L}_m$. 

The equations of motion given by the action in the Einstein representation are easily derived from (\ref{eq:actionER}). However, it is not trivial to figure out what is the source $\sigma$ of the scalar-field $\varphi$ in the Einstein representation, where $\sigma= (-\tilde{g})^{-1/2} \delta S_m/\delta {\varphi}$. For instance, \cite{DamPolyGRG94,DamPolyNPB94} use an assumption on the functional dependency of the Einstein mass $\ut{m}$  (\ref{eq:assumpt}); instead of deriving the dependency from the action in the original representation. In the following, we expound the derivation of $\sigma$. 

The variation of equation (\ref{eq:actionMJ}) for relevant fields leads to:
\begin{equation}
\delta S_m=\int d^4x\sqrt{-g} \left(-f(\Phi) T_{\alpha \beta} ~\delta g^{\alpha \beta}+2 f_{,\Phi}(\Phi) \mathcal{L}_m ~\delta \Phi \right).\label{eq:varSm}
\end{equation}
Now, since one has $g^{\alpha \beta} \equiv \Phi \tilde{g}^{\alpha \beta}$, the variation of the \textit{physical} metric gives
\begin{equation}
\delta g^{\alpha \beta}= \tilde{g}^{\alpha \beta} ~\delta \Phi + \Phi ~\delta \tilde{g}^{\alpha \beta}.
\end{equation}
Therefore, equation (\ref{eq:varSm}) writes:
\begin{eqnarray}
\delta S_m&=&\int d^4x\sqrt{-g} (-\Phi f(\Phi) T_{\alpha \beta} ~\delta \tilde{g}^{\alpha \beta} \\
&& + \left[-f(\Phi) \tilde{g}^{\alpha \beta} T_{\alpha \beta}+ 2 f_{,\Phi}(\Phi) \mathcal{L}_m \right]~ \delta \Phi ).\nonumber
\end{eqnarray}
Now, using $T_{\alpha \beta}= \Phi \tilde{T}_{\alpha \beta}$, $\tilde{T} \equiv \tilde{g}^{\alpha \beta} \tilde{T}_{\alpha \beta}$, $\delta \varphi = \delta \Phi / \Phi$ and $\Phi f_{,\Phi}(\Phi)=f_{,\varphi}(\Phi(\varphi))$, one gets:
\begin{eqnarray}
\delta S_m=&&\int d^4x\sqrt{-\tilde{g}} \big(- f(\Phi(\varphi)) \tilde{T}_{\alpha \beta}~ \delta \tilde{g}^{\alpha \beta} \label{eq:deltaSM} \\
&&- \left[1- 2 \frac{f_{,\varphi}(\Phi(\varphi))}{f(\Phi(\varphi))} \frac{\tilde{\mathcal{L}}_m}{\tilde{T}} \right]f(\Phi(\varphi)) ~\tilde{T}  ~\delta \varphi \big).\nonumber
\end{eqnarray}
The second part of the right hand side of (\ref{eq:deltaSM}) gives the sought-after $\sigma$. 

Now, since $\tilde{\mathcal{L}}_m=\Phi^{-2} \mathcal{L}_m$ and $\tilde{T}= \Phi^{-2} T$, $\tilde{\mathcal{L}}_m / \tilde{T}$ reduces to $\mathcal{L}_m / T=1+O(c^{-2})$ \cite{moi_hPRD12}. Note that for non-interacting particles ($P=0$), one has $\tilde{\mathcal{L}}_m / \tilde{T}=1$ \cite{moiPRD13}.

\subsection*{Rescaling of the scalar-field, and correction of Damour and Polyakov's equation for $\gamma$}
\label{sec:RSSF}

While one can work with the action (\ref{eq:actionER}) in the Einstein representation, the scalar-field is often rescaled such that the action writes:
\begin{eqnarray}
S= \int d^4x\sqrt{-\tilde{g}} \left(  \tilde{R} - \tilde{g}^{\alpha \beta} \partial_\alpha \ul{\varphi} \partial_\beta \ul{\varphi} \right) + S_m, \label{eq:SEwRS}
\end{eqnarray}
where $\d \ul{\varphi}=\pm \sqrt{\omega + 3/2} ~\d \varphi$. In what follows we consider the \textit{re-scaled} action (\ref{eq:SEwRS}) only in order to compare our result with previous studies. Choosing the re-scaling $\d \ul{\varphi}= \sqrt{\omega + 3/2} ~\d \varphi$, one can re-write (\ref{eq:deltaSM}) as follows:
\begin{eqnarray}
&&\delta S_m=\int d^4x\sqrt{-\tilde{g}} \big(- f(\Phi(\ul{\varphi})) \tilde{T}_{\alpha \beta}~ \delta \tilde{g}^{\alpha \beta} \label{eq:deltaSM2} \\
&&- \left[1- 2 \frac{f_{,\varphi}(\Phi(\varphi))}{f(\Phi(\varphi))} \frac{\tilde{\mathcal{L}}_m}{\tilde{T}} \right] \frac{f(\Phi(\ul{\varphi}))}{\sqrt{\omega(\Phi(\ul{\varphi}))+3/2}} ~\tilde{T}  ~\delta \ul{\varphi} \big).\nonumber
\end{eqnarray}
Note that the second part of the right hand side gives the source $\ul{\sigma}$ of the scalar-field $\ul{\varphi}$, with $\ul{\sigma}= (-\tilde{g})^{-1/2} \delta S_m/\delta \ul{\varphi}$. Also note that for the non-interacting particles treated in \cite{DamPolyGRG94,DamPolyNPB94} ($P=0$), one has $\tilde{\mathcal{L}}_m / \tilde{T}=1$ \cite{moiPRD13}. From (\ref{eq:SEwRS}) and (\ref{eq:deltaSM2}), one gets the following 1.5PN/RM \cite{moiCQG11} equation for $\ul{\varphi}$:
\begin{equation}
\triangle \ul{\varphi} = - \alpha_0 f(\Phi_0) (1+\Upsilon) \tilde{T}+O(c^{-4}), 
\end{equation}
with $\alpha$ defined in \cite{damourCQG92} by $\alpha \equiv \partial \ln{A}/ \partial \ul{\varphi}$ with $g_{\alpha \beta}= A^2(\ul{\varphi}) \tilde{g}_{\alpha \beta}=\Phi^{-1}\tilde{g}_{\alpha \beta}$. In order to compare with \cite{DamPolyGRG94,DamPolyNPB94}, let us write:
\begin{equation}
\triangle \ul{\varphi} =  - \alpha_2 ~f(\Phi_0)\tilde{T}+O(c^{-4}) , \label{eq:SFERresc}
\end{equation}
with 
\begin{equation}
\alpha_2 \equiv \alpha_0~ (1+\Upsilon). \label{eq:SFERrescalpha}
\end{equation}
Therefore, as suggested in section \ref{sec:string}, the coupling strength $\alpha_2$ is in general different from $\alpha_0=\partial \ln A / \partial \ul{\varphi}|_{\Phi_0}$. On the other hand, from (\ref{eq:SEwRS}) and (\ref{eq:deltaSM2}), the Newtonian potential in the Einstein representation satisfies:
\begin{equation}
c^{-2}~\triangle \tilde{w} = -  f(\Phi_0) \tilde{T}^{00}+O(c^{-4}). \label{eq:SFER}
\end{equation}
Therefore, one has $g_{\alpha \beta}=A^2~\tilde{g}_{\alpha \beta}= (1-2c^{-2}~ \alpha_0 \alpha_2~ \tilde{w}+O(c^{-4}))~ \tilde{g}_{\alpha \beta}$  \footnote{see section \ref{sec:string}.}. Now, remembering that the Einstein metric satisfies the strong spatial isotropy condition ($\tilde{g}_{ij} \tilde{g}_{00} = - \delta_{ij} + O(c^{-4})$), one gets for the PN parameter $\gamma$:
\begin{equation}
\gamma = \frac{1- \alpha_0 \alpha_2}{1+ \alpha_0  \alpha_2} =\frac{1- \alpha_0^2 (1+\Upsilon)}{1+ \alpha_0^2 (1+\Upsilon)}, \label{eq:gammaER}
\end{equation}
or
\begin{equation}
\ul{\gamma} = -\frac{2 \alpha_0^2 (1+\Upsilon)}{1+ \alpha_0^2 (1+\Upsilon)},
\end{equation}
where $\ul{\gamma} \equiv \gamma-1$.  Therefore, it shows that the parameter $\Upsilon$ is missing in the formula for the $\gamma$ parameter given in \cite{DamPolyGRG94,DamPolyNPB94,damourPRD10,damourPRL02,damourPRD02}. Accordingly, their parameter can only be less than one (see (9) in \cite{DamPolyGRG94} for instance); while we have shown that, depending on the coupling function, it could actually be either positive, null or negative. 

The discrepancy comes from the wrong assumption in \cite{DamPolyGRG94,DamPolyNPB94} that $\partial \ln A / \partial \ul{\varphi} = \partial \ln \ut{m} / \partial \ul{\varphi}$. In (\ref{eq:SFERrescalpha})  we show how the coupling strength $\alpha_2$ defined in (\ref{eq:phiA2}) relates to $\alpha_0=\partial \ln A / \partial \ul{\varphi}|_{\Phi_0}$ in general.

Now, remembering that $\alpha_0^{-2} = 2\omega_0+3$, one exactly gets (\ref{eq:gammadef}) from (\ref{eq:gammaER}). 

\begin{acknowledgments}
This research was partly supported by an appointment to the NASA Postdoctoral Program at the Jet Propulsion Laboratory, California Institute of Technology, administered by Oak Ridge Associated Universities through a contract with NASA. \copyright 2012 California Institute of Technology. Government sponsorship acknowledged. \\
This research was partly done as an invited researcher of the Observatoire de la C\^ote d'Azur.\\
The author wants to thank Aurelien Hees, Tiberiu Harko, Francisco Lobo, John Moffat and Viktor Toth for their interesting comments.\\

\end{acknowledgments}

\end{document}